\documentclass{mn2e}

\usepackage{times}
\usepackage{rotating}
\usepackage{xtab}
\usepackage{psfig}

\newif\ifAMStwofonts
\AMStwofontstrue

\def\til{$\sim$\thinspace}

\def\deg{$^{\circ}$}
\def\six{${\omega}_{_{650}}$}
\def\thr{${\omega}_{_{370}}$}
\def\two{${\omega}_{_{230}}$}
\def\ssix{${\omega}_{_{650}}$\space}
\def\sthr{${\omega}_{_{370}}$\space}
\def\stwo{${\omega}_{_{230}}$\space}
\def\th{\thinspace}

\title[The Pulsating Primary of the CV GW Librae]
{The Nonradially-Pulsating Primary of the Cataclysmic Variable GW Librae}
\author[L. van Zyl et al.]
       {L. van Zyl$^{1,3}$\thanks{E-mail: lvz@astro.keele.ac.uk}, B. Warner$^{3}$, D. O'Donoghue$^{4}$, C. Hellier$^{1}$, P. Woudt$^{3}$, D. Sullivan$^{5}$,
\and
J. Pritchard$^{10}$, J. Kemp$^{6}$, J. Patterson$^{6}$, W. Welsh$^{7,11}$, J. Casares$^{8}$, T. Shahbaz$^{2,8}$,\and
F. van der Hooft$^{12}$, S. Vennes$^{9}$\\
\\
       $^{ 1}$Astrophysics Group, School of Chemistry and Physics,
              Keele University, Staffordshire, ST5 5BG\\
       $^{ 2}$Department of Astrophysics, Oxford University, Keble Road, 
              Oxford OX1 3RH\\
       $^{ 3}$Department of Astronomy, University of Cape Town, Rondebosch 7700,
              South Africa\\
       $^{ 4}$South African Astronomical Observatory, PO Box 9,
              Observatory 7935, South Africa\\
       $^{ 5}$School of Chemical and Physical Sciences, Victoria University, 
              Box 600, Wellington, New Zealand\\
       $^{ 6}$Department of Astronomy, Columbia University, 
              550 West 120th Street, New York, NY 10027, USA\\
       $^{ 7}$Department of Astronomy and McDonald Observatory, University of Texas at Austin, Austin, TX 78712, USA\\
       $^{ 8}$Instituto de Astrof\'{i}sica de Canarias,
              38200 La Laguna, Tenerife,Spain\\
       $^{ 9}$Department of Mathematics and Astrophysical Theory Centre,
              Australian National University, Canberra, ACT 0200, Australia\\
       $^{10}$Department of Physics and Astronomy, University of Canterbury,
              Christchurch, New Zealand\\
       $^{11}$Department of Astronomy, San Diego State University, 5500 Campanile Drive, San Diego, CA 92182-1221, USA\\
       $^{12}$Astronomical Institute `Anton Pannekoek', University of Amsterdam; and Center               for High Energy Astrophysics,\\
              Kruislaan 403, NL-1098 SJ Amsterdam, The Netherlands
       }
\date{Accepted 
      Received }

\pagerange{\pageref{firstpage}--\pageref{lastpage}}
\pubyear{2001}

\begin{document}

\maketitle

\label{firstpage}

\begin{abstract}
The dwarf nova GW Librae is the first cataclysmic variable discovered to have a primary in a white dwarf instability strip, making it the first multi-mode, nonradially-pulsating star known to be accreting. The primaries of CVs, embedded in hot, bright accretion discs, are difficult to study directly. Applying the techniques of asteroseismology to GW Librae could therefore give us an unprecedented look at a white dwarf that has undergone $\sim 10^9$ years of accretion. However, an accreting white dwarf may have characteristics sufficiently different from those of single pulsating white dwarfs to render the standard models of white dwarf pulsations invalid for its study. This paper presents amplitude spectra of GW Lib from a series of observing campaigns conducted during 1997, 1998 and 2001. We find that the dominant pulsation modes cluster at periods near 650, 370 and 230~s, which also appear in linear combinations with each other. GW Lib's pulsation spectrum is highly unstable on time-scales of months, and exhibits clusters of signals very closely spaced in frequency, with separations on the order of 1~$\mu$Hz.
\end{abstract}


\section{Introduction}


\begin{figure}
\psfig{file=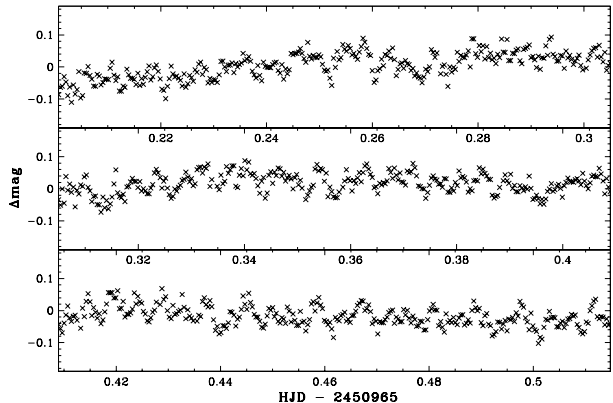,width=8.0cm}
\caption{A sample of GW Lib's light-curve, obtained with the SAAO 0.75-m telescope on 31/05/1998.}\label{fig:lcv}
\end{figure}

The dwarf nova (DN) GW Librae was the first cataclysmic variable (CV) discovered to contain a nonradially pulsating white dwarf (DAV, or ZZ Ceti star; van Zyl et al. 2000). The existence of a DAV within an interacting binary system has some potentially significant implications for the study of accreting stars. White dwarfs in CV systems, with their long accretion histories and many novae eruptions during their $\sim10^9$-$10^{10}$-year lifetimes, may have very different internal structures from single white dwarfs (Townsley \& Bildsten 2002). However, the primaries of CVs are difficult to study: the flux from CVs tends to be dominated by accretion processes. Nonradial pulsations, which carry a wealth of information about stellar structure, composition and evolution (e.g., Winget et al. 1991, 1994), therefore potentially provide us with a unique and very powerful tool with which to study CV evolution and the accretion process.

In addition, the existence of an accreting DAV also has some exciting implications for the study of single white dwarfs. As DA white dwarfs cool on evolutionary time-scales, taking 5-$10 \times 10^8$ years to pass through the DAV instability strip (Wood 1992), their pulsations change in ways not yet fully understood, becoming more unstable and prone to nonlinear effects. In the case of a DAV in a CV system which undergoes periodic DN outbursts, it may be possible to observe the star move through the traditional DAV instability strip in a matter of only months or years, \til$10^8$ times more rapidly than single DAVs, as the white dwarf photosphere cools between DN outbursts.


\subsection{The first accreting multi-mode pulsating star}

GW Librae is a faint DN which called attention to itself for the first time in 1983 when its brightness increased by $\sim$ 9 mag (Gonz\'{a}lez 1983) -- its only observed outburst. The large amplitude of the outburst initially led to GW Lib being misclassified as a nova; however, subsequent spectroscopy (Duerbeck \& Sitter 1987; Ringwald, Naylor \& Mukai 1996) showed it to be a system with a very low mass-transfer rate. GW Lib is almost certainly a member of the class of long outburst-interval, low mass-transfer-rate DNe known as the WZ Sagittae stars.

GW Lib's faintness (V$\sim$18) precluded interest in it for 14 years following its outburst, until we chanced upon its remarkable properties while conducting a high-speed-photometry survey of faint Southern Hemisphere CVs. Its spectrum and its 1983 WZ-Sge-style superoutburst show beyond doubt that GW Lib is a CV, but the Fourier transform of its light-curve (Fig.~\ref{fig:lcv}) resembled that of a nonradially pulsating single white dwarf. Nonradial pulsations had never been observed in an accreting white dwarf before (except possibly a single mode in the DOV primary of the CV AM CVn [Solheim et al. 1998]); it had been assumed either that accretion would keep the primaries of CVs too hot to pulsate, or that He in the accreted material would suppress pulsations. The DAV (or ZZ Cet) instability strip for (non-accreting) hydrogen-atmosphere white dwarfs (DAs) occurs between T$_{\rm eff}$s of \til 11,000 and \til 12,500~K (Koester \& Holberg 2001), depending on the mixing length prescription used to describe convection. The coolest DN primaries have surface temperatures of typically 15~000~K (Sion, Urban \& Lyons 2001).

A spectroscopic study of GW Lib by Thorstensen et al. (2002) found an orbital period of 76.78~min -- the shortest known orbital period for a CV with a hydrogen-rich donor (V485 Cen has a period of 59~min, but its companion has a low hydrogen content [Augusteijn et al. 1996]). This orbital period supports GW Lib's status as a WZ Sge star. The WZ Sge stars, or `TOADs', are the oldest DNe, and are characterized by very low mass-transfer rates, long outburst recurrent times and short orbital periods (e.g. Bailey 1979; O'Donoghue et al. 1991; Howell, Szkody \& Cannizzo 1995). Along with the magnetic AM Her CVs, which also have low mass-transfer rates, WZ Sge stars are therefore the most likely CVs to harbour DAVs, because of their age (white dwarfs take 0.5--5 Gyr to cool to 12,500~K, depending on their mass [Chabrier et al. 2000]), and because of their low mass-transfer rate (so that compressional heating of the white dwarf's cores through accretion is small [Townsley \& Bildsten 2001]).

Despite extensive photometry, no orbital modulation has been detected in GW Lib's light-curves, suggesting that the system has a low inclination. From the width of GW Lib's emission lines, Thorstensen et al. (2002) infer an inclination of \til 11\deg. Thorstensen et al. also find an absolute magnitude for GW Lib of $M_V$ \til 11.5, a distance of \til 125~pc and a significant proper motion of $66 \pm 12$~mas~y$^{-1}$.

Szkody, Desai \& Hoard (2000) found that spectra of GW Lib were roughly consistent with an effective temperature of 11,000$\pm$1,000~K (assuming that 100\% of the flux came from the white dwarf), which put GW Lib in the DAV instability strip. However, Szkody et al. (2002) then obtained a more accurate white dwarf surface temperature, with HST UV spectroscopy, of 14,700~K, which puts it well outside the traditional DAV instability strip. This may imply a possible temporary heating of the white dwarf atmosphere due to an unobserved outburst (while the core and mean temperatures are much lower), or that the instability strip for accreting DAVs is different from that of single DAVs. Indeed, the notion of an `instability strip' for accreting white dwarfs may be meaningless: instability strips are usually associated with groups of stars moving by single star evolution through a T$_{\rm eff}$ interval, but in accreting systems the mass transfer onto the white dwarf would very likely introduce very important additional parameters.

GW Lib's pulsation spectrum shows the unstable behaviour typical of the cool, large amplitude DA white dwarf pulsators. As white dwarfs evolve from the hot to the cool end of the DAV instability strip, their pulsation spectra become increasingly more complex and unstable, the amplitudes of their pulsation modes often changing dramatically from month to month (Kleinman et al. 1998). The amplitudes of the oscillations in GW Lib's light-curve are very small ($\sim$ 5-10~mmag). However, as we expect approximately half the light from the system to be coming from the accretion disc (based on a comparison with the DN Z Cha), the intrinsic amplitude of the pulsations is likely to be greater.

\subsection{Problems specific to accreting pulsators}

As a result of its birth in a common envelope and its subsequent long history of accretion, it would not be surprising if GW Lib's primary were to have characteristics, and therefore a pulsation spectrum, different from those of single white dwarfs. We discuss below some issues which we believe may be of significance in the case of an accreting DAV.

\subsubsection{Accretion-induced spherical asymmetries}

If a CV primary has a weak or absent magnetic field (as is the case of GW Lib, which does not appear as an X-ray source in the ROSAT All-Sky Survey Faint Source Catalogue [Voges et al. 2000]), material from the disc will accrete onto the equatorial regions of the star. The white dwarf therefore develops an equatorial band which has a different temperature, rotational and chemical composition structure to the cooler, higher-latitude regions of the star. The effect is especially pronounced after superoutbursts. Sparks et al. (1993) find that after WZ Sge's superoutbursts, the hot accreted material is confined to a broad equatorial band on the primary for $\sim$10 years. Nonradial modes are sensitive to such latitudinal differences within the stellar envelope (e.g. Schou et al. 1998). However, single DAVs are spherically symmetric to a very high degree, and therefore their pulsation spectra have been free of these complexities.

\begin{figure}
\psfig{file=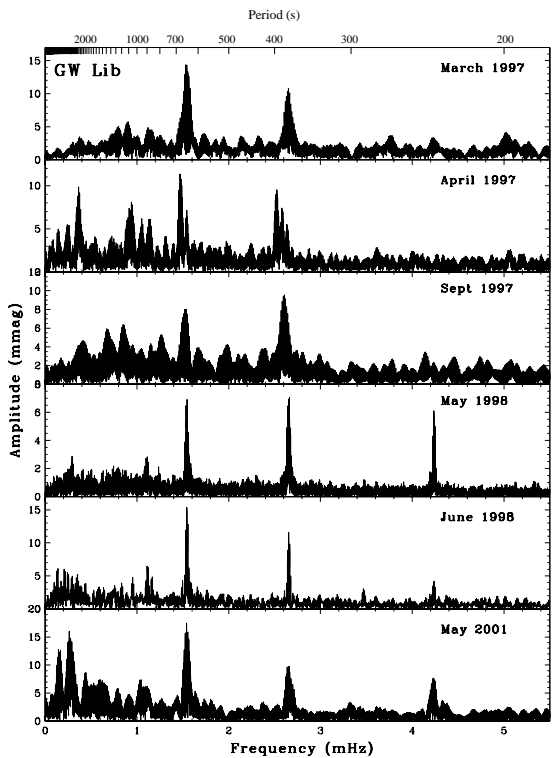,width=8cm}
\caption{The amplitude spectra of GW Lib from 6 observing campaigns from March 1997 to May 2001. The light-curves were prepared by fitting and removing linear and low-frequency trends.}\label{fig:full}
\end{figure}

\begin{figure*}
\hspace{0.1cm}
\psfig{file=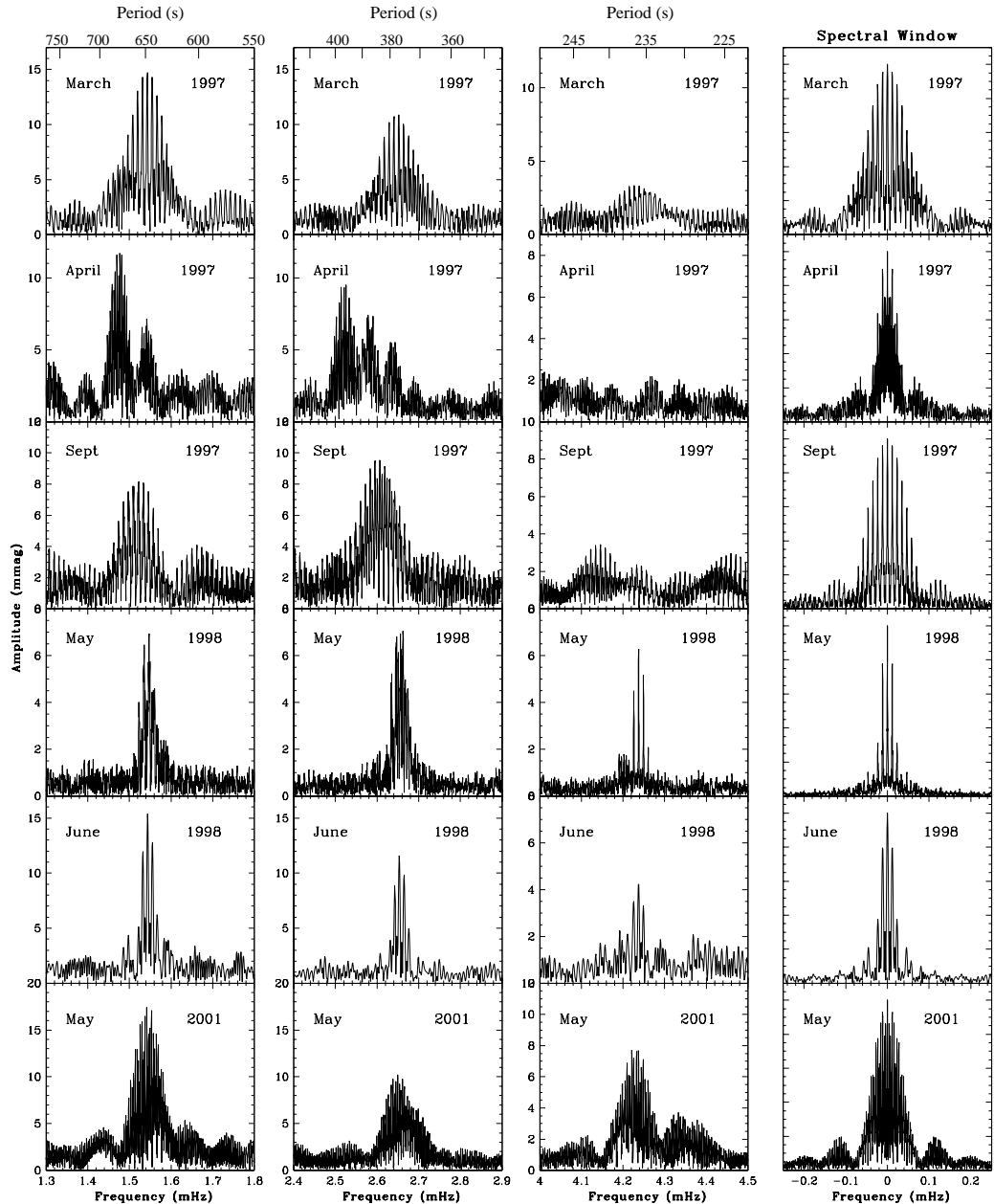,width=14cm}
\caption{Expanded plots of the 650-s, 370-s and 230-s regions of GW Lib's amplitude spectra. The fourth column gives the spectral windows for each observing campaign. The unstable nature of GW Lib's pulsations is clearly apparent. Our best data, with the highest signal-to-noise, the longest run-lengths and the cleanest spectral window, are the May 1998 observations. However, in May 1998 GW Lib's pulsation spectrum, apart from a new mode near 230~s, showed fewer active modes than in April 1997.}\label{fig:details}
\end{figure*}

As yet, no-one has tried to model the pulsational behaviour of an accreting DAV, so at present we can only speculate as to the effects of accretion on its pulsation spectrum. If accretion-induced spherical asymmetries extend to the depth of the pulsation driving zone, the pulsation eigenfunctions could be dramatically affected. 

In addition, it is possible that an accreting DAV may exhibit pulsation modes with $l>2$: in single DAVs, geometric cancellation effects result in modes with $l>2$ having amplitudes too low to be detected. However, an accreting DAV would have a different surface temperature distribution, which could alter the surface-temperature geometry of modes with $l>2$ in such a way that they may become observable.

\subsubsection{Rapid Rotation}

Single DAVs rotate very slowly, with periods on the order of several hours or longer (e.g. Koester et al. 1998, Bradley 2001), and their pulsations have periods on the order of hundreds of seconds. Therefore 
\begin{equation}
\Omega \ll {\omega}_{_{(k,l,m)}},
\end{equation}
where ${\omega}_{_{(k,l,m)}}$ are the pulsation eigenfrequencies and $\Omega$ is the stellar rotation frequency, and the leading effects of rotation on the oscillations are linear in $\Omega$. The effect of slow rotation is to lift the degeneracy of the ${\omega}_{_{(k,l,m)}}$ eigenfrequencies, splitting them into multiplets, where $m=0,\pm 1,...,\pm l$.

CV primaries, however, typically have rotation periods much shorter than those of single white dwarfs, because they are spun up by the accretion process (e.g. Warner 1995). The rotation periods for primaries in non-magnetic CVs are on the order of 100~s (Warner [1995] and references therein). In this case,
\begin{equation}
\Omega \sim {\omega}_{_{(k,l,m)}},
\end{equation}
and rotation can no longer be dealt with as a small perturbation, and therefore the pulsation modes can no longer be described by a single spherical harmonic $Y^l_m(\theta,\phi)$. Coriolis forces become of the same order as buoyancy forces which provide the restoring forces for $g$-modes, and therefore become a new restoring force for waves of a very different physical nature, which can also be trapped in the star and form global modes (Vorontsov 1993). The geometry of the Coriolis forces is significantly more complex, making the classification of the new modes difficult.


\begin{table}
\begin{center}
\caption{Summary of Observations}
\begin{tabular}{lccrr}\hline\hline
               & Start & Int. & Run-$\;\,$ & \\
$\;\;\;\;$Date & HJD   & time & Length & Telescope$^\star$ \\		          
 & (-2450000) & (s) & (h)$\;\;\,$      &  \\
 \hline
13/03/97 & 0521.58710 &  6 &  1.998 & {\sc saao 1.0}{\thinspace}m   \\
14/03/97 & 0522.49980 & 12 &  3.829 & {\sc saao 1.0}{\thinspace}m  \\
15/03/97 & 0523.56415 & 12 &  2.482 & {\sc saao 1.0}{\thinspace}m  \\
16/03/97 & 0524.55071 & 12 &  2.419 & {\sc saao 1.0}{\thinspace}m  \\
& & & & \\		          
01/04/97 & 0540.50686 & 20 &  1.737 & {\sc saao 1.0}{\thinspace}m  \\
02/04/97 & 0541.39409 & 20 &  6.509 & {\sc saao 1.0}{\thinspace}m  \\
04/04/97 & 0543.40310 & 20 &  4.558 & {\sc saao 1.0}{\thinspace}m    \\
07/04/97 & 0546.39368 & 20 &  5.969 & {\sc saao 1.0}{\thinspace}m  \\
& & & & \\		          
31/08/97 & 0692.27492 & 20 &  2.188 & {\sc saao 1.0}{\thinspace}m  \\
01/09/97 & 0693.22053 & 20 &  2.989 & {\sc saao 1.0}{\thinspace}m   \\
04/09/97 & 0696.22408 & 10 &  3.466 & {\sc saao 1.9}{\thinspace}m  \\
05/09/97 & 0697.21810 & 10 &  3.530 & {\sc saao 1.9}{\thinspace}m  \\
06/09/97 & 0698.22028 & 10 &  3.246 & {\sc saao 1.9}{\thinspace}m  \\
07/09/97 & 0699.22226 & 10 &  2.826 & {\sc saao 1.9}{\thinspace}m   \\
08/09/97 & 0700.21981 & 10 &  2.658 & {\sc saao 1.9}{\thinspace}m    \\
& & & & \\	      	          
19/05/98$^\dagger$ & 0953.28861 & 10 &  5.859 & {\sc saao 1.9}{\thinspace}m \\
20/05/98$^\dagger$ & 0954.30589 & 10 &  7.444 & {\sc saao 1.9}{\thinspace}m \\
21/05/98$^\dagger$ & 0955.28927 & 10 &  9.131 & {\sc saao 1.9}{\thinspace}m \\
22/05/98$^\dagger$ & 0956.34921 & 10 &  7.310 & {\sc saao 1.9}{\thinspace}m \\
23/05/98$^\dagger$ & 0956.92086 & 40 &  4.013 & {\sc mtj 1.0}{\thinspace}m \\
23/05/98$^\dagger$ & 0957.24252 & 10 & 10.211 & {\sc saao 1.9}{\thinspace}m \\
24/05/98$^\dagger$ & 0957.93488 & 40 &  5.378 & {\sc mtj 1.0}{\thinspace}m \\
24/05/98$^\dagger$ & 0958.23836 & 10 & 10.141 & {\sc saao 1.9}{\thinspace}m \\
25/05/98$^\dagger$ & 0959.23145 & 20 & 10.316 & {\sc saao 1.9}{\thinspace}m \\
26/05/98$^\dagger$ & 0960.31582 & 30 &  6.400 & {\sc saao .75}{\thinspace}m \\
28/05/98$^\dagger$ & 0962.35375 & 30 &  5.890 & {\sc saao .75}{\thinspace}m  \\
29/05/98$^\dagger$ & 0963.24397 & 30 &  8.502 & {\sc saao .75}{\thinspace}m   \\
30/05/98$^\dagger$ & 0963.91411 & 60 &  7.300 & {\sc mssso 1.0}{\thinspace}m \\
30/05/98$^\dagger$ & 0964.20302 & 30 &  9.541 & {\sc saao .75}{\thinspace}m \\
31/05/98$^\dagger$ & 0965.20068 & 30 &  9.274 & {\sc saao .75}{\thinspace}m \\
01/06/98$^\dagger$ & 0966.23016 & 30 &  3.296 & {\sc saao .75}{\thinspace}m  \\
& & & & \\
16/06/98$^\ddagger$ & 0981.22138 & 30 &  8.520 & {\sc saao .75}{\thinspace}m  \\
17/06/98$^\ddagger$ & 0982.20179 & 30 &  8.754 & {\sc saao .75}{\thinspace}m  \\
18/06/98$^\ddagger$ & 0983.24030 & 30 &  8.223 & {\sc saao .75}{\thinspace}m  \\
& & & & \\
27/06/98 & 0992.63858 & 20 & 1.128 & {\sc mcd 1.2}{\thinspace}m \\
28/06/98 & 0993.64537 & 20 & 0.836 & {\sc mcd 1.2}{\thinspace}m \\
29/06/98 & 0994.63663 & 20 & 1.080 & {\sc mcd 1.2}{\thinspace}m \\
& & & & \\
18/05/01 & 2048.40493 & 20 & 4.88 & {\sc saao 1.0}{\thinspace}m \\
20/05/01 & 2050.54336 & 20 & 2.49 & {\sc saao 1.0}{\thinspace}m \\
21/05/01 & 2051.27101 & 20 & 4.98 & {\sc saao 1.0}{\thinspace}m \\
23/05/01 & 2053.40757 & 15 & 3.51 & {\sc saao 1.9}{\thinspace}m \\
26/05/01 & 2056.41459 & 20 & 5.29 & {\sc saao 1.9}{\thinspace}m \\
\hline \hline
\end{tabular}
\begin{tabular}{p{0.01cm}p{7.1cm}}
$^\star$& {\footnotesize {\sc saao} = South African Astronomical Observatory, {\sc mtj} = Mt John Observatory (New Zealand), {\sc mssso} = Mt Stromlo \& Siding Spring Observatory (Australia), {\sc mcd} = McDonald Observatory (USA).}\\
$^\dagger$ & {\footnotesize The May 1998 observations are our best data, with the longest run-lengths, highest signal-to-noise and cleanest spectral windows.}\\
$^\ddagger$ & {\footnotesize These observations from June 1998 are our next best dataset, and when combined with the May 1998 dataset to extend its baseline, make up the `May\&June98' dataset referred to in the text.}
\end{tabular}
\end{center}
\end{table}

\subsubsection{Non-stationary pulsation spectra}

During a superoutburst, the temperature of a WZ Sge primary's photosphere can increase by a factor of 2 or 3, and take \til10 years to return to its quiescent value (Sion \& Szkody 1990; Sparks et al. 1992; Long et al. 1994). The characteristics of DAV pulsations are sensitive to temperature, and therefore superoutbursts are likely to have an effect on the pulsation spectrum of a CV-primary DAV. Frequency and amplitude modulation as a result of temperature-induced changes in the envelope could make the pulsation spectrum very difficult to unravel.

In addition to steady temperature changes of the photosphere resulting from superoutbursts, there may also be shorter-timescale accretion-induced temperature perturbations which may affect the pulsation modes. Flickering in DNe light-curves indicates that accretion is very `blobby' (Warner 1995). Blobs of material accreting onto a star unstable to pulsations may behave like a stochastic hammer applied to an oscillator, inducing constant phase changes in the oscillations, which would also complicate the pulsation spectrum.


\section{The pulsation spectrum of GW Lib}

\subsection{The observations}

We conducted six observing campaigns on GW Lib: in March, April and September 1997, May and June 1998, and May 2001. A summary of the observations is given in Table 1. The bulk of the data were obtained from the South African Astronomical Observatory (SAAO) in Sutherland, South Africa, using a high-speed, frame-transfer CCD photometer purpose-built for observing blue, rapidly-varying stars like pulsating white dwarfs and CVs (O'Donoghue 1995). A large part of the data we discuss has appeared in the proceedings of the Fifth Whole Earth Telescope Workshop (van Zyl et al. 2000).

In order to attempt to resolve GW Lib's pulsation spectrum, we conducted a multi-longitude observing campaign in May 1998. The participating observatories were Mt John Observatory (New Zealand), Mt Stromlo \& Siding Spring Observatory (Australia), CTIO (Chile) and SAAO. All sites made use of CCDs, and observations were made in white light (due to GW Lib's faintness), but only SAAO was able to use a blue-sensitive, back-illuminated chip. Very poor weather at all the sites apart from SAAO resulted in us obtaining only two half-nights of data from Mt John, only one night from Mt Stromlo, and no usable observations from CTIO. From SAAO we were able to observe on 13 of 14 nights, and on most of these nights were able to obtain good quality, long light-curves.

In order to extend the baseline of the May 1998 dataset (and thereby increase the resolution of its Fourier transform), we obtained 3 good runs at SAAO on 3 successive nights in June 1998, separated from the May 1998 data by 14 days. We also obtained short light-curves from an additional 3 nights in June 1998 from McDonald Observatory (USA), 9 days after the end of the June SAAO run. However, adding these data to the SAAO dataset complicated the spectral window to an extent which negated any benefit from further extending the baseline. Therefore, in what follows, `May98' refers to the May 1998 multisite campaign, and `May\&June98' refers to the May98 dataset with its baseline extended with the inclusion of the 3 nights of SAAO data obtained 14 days later.

\begin{figure}
\psfig{file=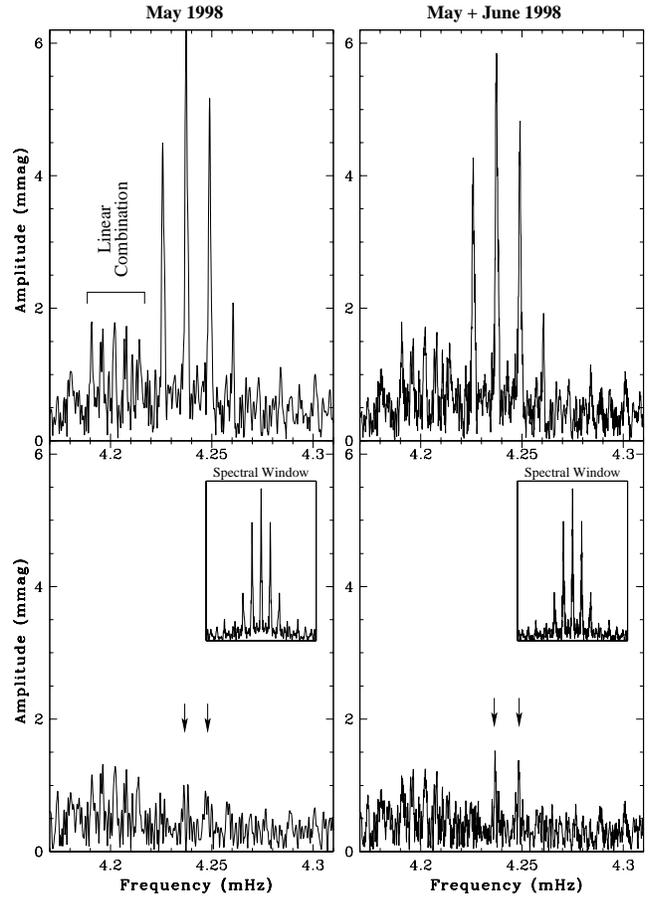,width=8.4cm}
\caption{The 230-s mode. In the amplitude spectrum of the May98 data alone, this feature appears to be a single mode (left). However, with the addition of 3 nights in June 1998, the baseline is sufficiently extended to reveal a second, low-amplitude mode after prewhitening with the dominant mode (right). (The `bump' left of centre is a linear combination mode.)}\label{fig:mode236}
\vspace{4mm}
\end{figure}

\begin{figure}
\psfig{file=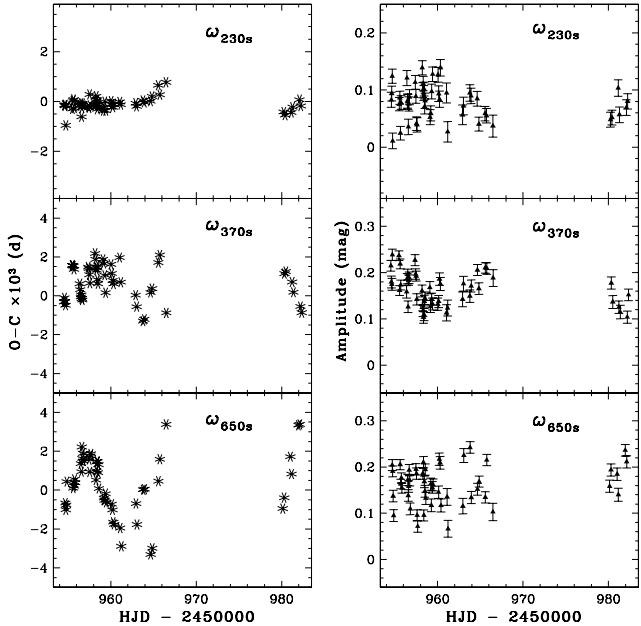,width=8.4cm}
\caption{O-C diagrams (left) and amplitudes (right) of the three principal structures,\two, \sthr and \six, in the May\&June98 amplitude spectrum. The points give the O-C and amplitude of the structures for the first and second half of each night, calculated using the mean frequency of each structure's components from Table 2.}\label{fig:ominusc}
\end{figure}

\subsection{The amplitude spectra}

The amplitude spectra presented in this paper are Fourier transforms (FTs) of GW Lib's light-curves, obtained using an algorithm for calculating discrete FTs of unequally-spaced data proposed by Deeming (1975), and with a short-cut proposed by Kurtz (1985). In Fig.~\ref{fig:full} we present amplitude spectra of GW Lib from each of the six observing campaigns. Fig.~\ref{fig:details} shows the regions of interest in the amplitude spectra in greater detail.


GW Lib's pulsation spectrum is at its most active in April 1997, when there appear to be two modes active near 650~s and three near 370~s (Figs.~\ref{fig:full} and {fig:details}). The structure of the modes in the April 1997 Fourier transform is a great deal more complex than the window function (Fig.~\ref{fig:details}), indicating that each mode consists of more than one component. However, the April 1997 dataset does not have the baseline, nor the duty cycle, necessary to resolve the fine structure. 

By the time of the May98 multisite campaign, however, GW Lib's pulsation spectrum had changed dramatically. Three of the five pulsation modes present near 650~s and 370~s in April 1997 had disappeared by May 1998, and a new one near 230~s, not previously seen, had become detectable. For convenience, we shall refer to the three principal structures in the May98 and May\&June98 amplitude spectrum, at 650~s, 370~s and 230~s, as \six, \sthr and \two. However, we show below that each of these is a multiplet consisting of more than one frequency. 

In Table 2 we list all the signals present in each amplitude spectrum. Each signal consists of a set of aliases (ambiguities caused by gaps in the data), and Table 2 lists only the highest-amplitude alias for each signal. There is ambiguity in the aliases (the highest-amplitude alias is not necessarily the true frequency), so each frequency $\omega $ in Table 2 should be read as a set of frequencies $\omega + n \times 11.57 \mu$Hz, where $n=\pm1,\pm2,\pm3, ...$, and 11.57\th$\mu$Hz is the dominant 1-day alias.

The procedure for determining the amplitudes and frequencies of the signals in the amplitude spectrum is as follows. First, the mean and second-order trends (due to sky transparency) are removed from the light-curves, and then the amplitude spectrum is calculated. Then, assuming there are no aliasing ambiguities, the frequency of the highest peak is fitted to the data and subtracted (a process known as `pre-whitening'). The amplitude spectrum of the residuals is then computed and the process repeated until the data are consistent with noise. The frequencies found in this way (linear combinations as well as principal frequencies) are then fitted simultaneously to the full data set by a non-linear least-squares fit, which produces our final set of frequencies and amplitudes.



\subsection{The mode at 230~s}

\begin{figure*}
\hspace{1mm}\psfig{file=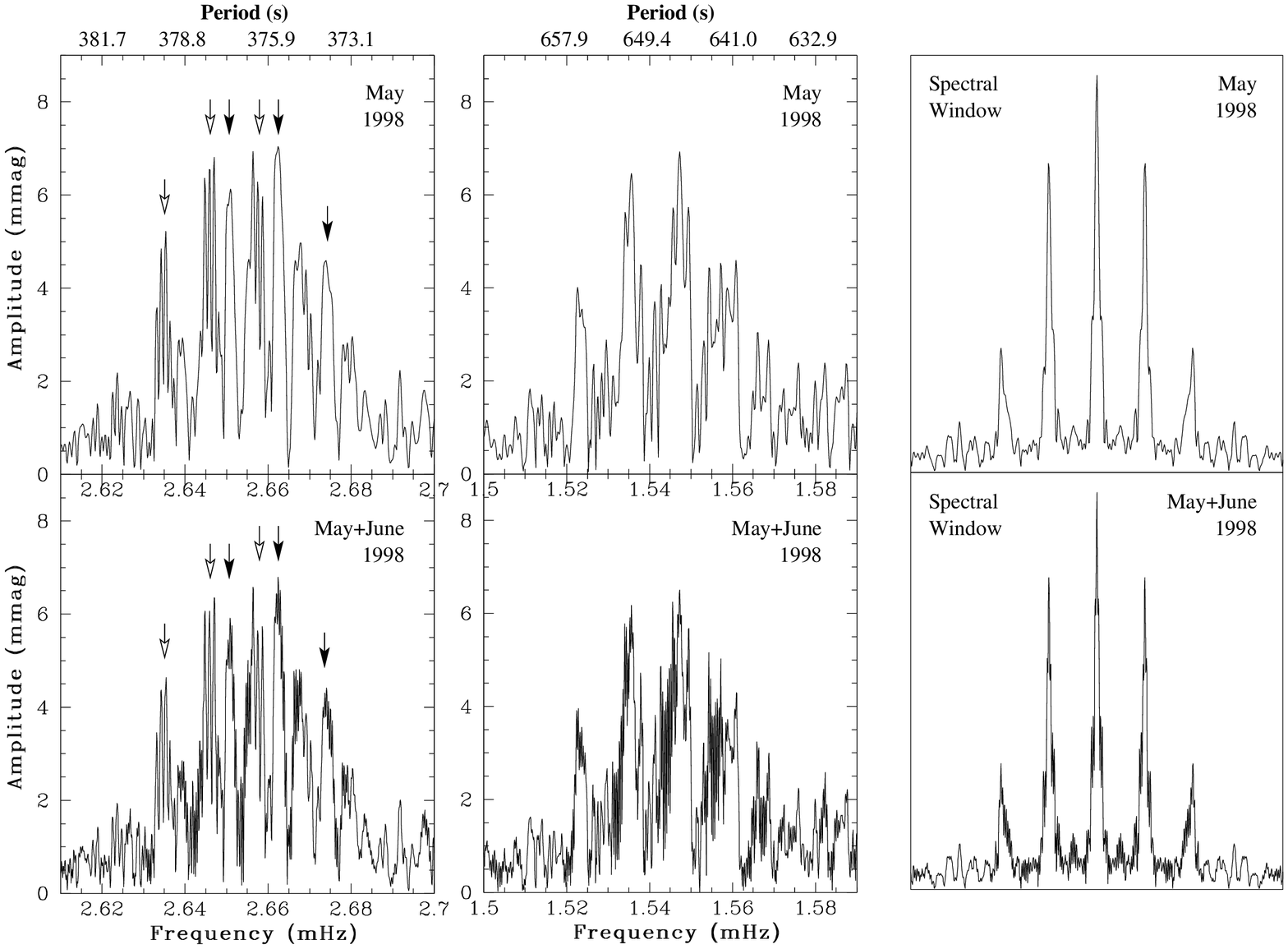,width=16cm}
\caption{Details of the amplitude spectra of the May98 (upper panels) and May\&June98 datasets (lower panels). The power in these regions is dominated by very fine frequency splitting, on the order of \til1~$\mu$Hz, not all of which can be explained in terms of artifacts generated by the window function. The `mode' we call \sthr (left-hand panels) appears to consist of two triplets, each with components spaced \til1~$\mu$Hz apart. One triplet (the three highest-amplitude aliases indicated by the open arrows) is centred at 2645.90~$\mu$Hz or 377.94~s. The second structure (the three highest-amplitude aliases indicated by the solid arrows) appears to be an unresolved triplet centred at 2662.94~$\mu$Hz or 375.52~s.
}\label{fig:yukky}
\vspace{4mm}
\end{figure*}

Ironically, the simplest feature of the May98 amplitude spectrum is also the most obviously unstable: there is no trace of the mode near 230~s in any of the 1997 amplitude spectra, although it is strong in 2001. At first glance, \stwo (Fig.~\ref{fig:mode236}, upper panels) appears identical to the spectral window in both the May98 and the May\&June98 FTs. The lower panels of Fig.~\ref{fig:mode236} show the FTs after prewhitening with this signal. In the May\&June98 spectrum, with its extended baseline, an additional, low-amplitude component is revealed. 

If the 230-s signal is a rotationally-split doublet (the most common reason for splitting of DAV modes), the frequency difference between the components, 0.79~$\mu$Hz, is remarkably small. If this is an $l=1$ mode (the most common modes in DAVs), the implied rotation period of the star is 7.3 days (from the properties of pulsational $g$-modes; see for example Winget et al. [1991,1994]). This rotation period is much longer than what we would expect for a CV primary.

A splitting of 0.79~$\mu$Hz corresponds to a beat period of \til15 days, which is very close to both the baseline of the May98 dataset, and the gap between the May and June data.  However, there are no splittings equal to 0.79~$\mu$Hz in the spectral window. Further evidence that \two\ is a closely-spaced doublet comes from changes in its O-C phases and amplitude over the observing campaign. In Fig.~\ref{fig:ominusc} we show the O-C and amplitudes of the three principal regions of power, \two, \sthr and \six, over the baseline of the May\&June98 dataset. The points were calculated using the dominant signal in \two\ and an average of the frequencies for each of \sthr and \ssix listed in Table 2. From the top panels in Fig.~\ref{fig:ominusc} it can be seen that the O-C curve of \stwo is not flat (Fig.~\ref{fig:ominusc}), which could indicate that \stwo consists of two components. Fig.~\ref{fig:ominusc} also shows that the amplitude of \stwo appears to be varying sinusoidally, which could indicate the existence of two closely-spaced components beating against each other.

\begin{sidewaystable*} 
\begin{tabular}{p{23.5cm}}
{\bf Table 2.} $\;$ The major periodicities (excluding linear combinations) detected in GW Lib's amplitude spectra from 1997 to 2001. Low-frequency signals have been included, but the list is likely to be incomplete as low-frequency trends were removed from the light-curves in order to bring down the level of the noise. \\
\end{tabular}
\begin{tabular}{rrrrrrrrrrrrrrrrrrrrrrrrr}
                    \hline 
                     \hline 
      &       &      & &        &         &      & &        &         &      &&          &      & &       &          &      & &         &      & &       &      &      \\ 
\multicolumn{3}{c}{\large {\bf Mar 1997}} & & \multicolumn{3}{c}{\large {\bf Apr 1997}} & & \multicolumn{3}{c}{\large {\bf Sept 1997}} & & \multicolumn{2}{c}{\large {\bf May 1998}} & & \multicolumn{3}{c}{\large {\bf May+June 98}} && \multicolumn{2}{c}{\large {\bf June 1998}} & & \multicolumn{3}{c}{\large {\bf May 2001}} \\
      &       &      & &        &         &      & &        &         &      &&          &      & &       &          &      & &         &      & &       &      &      \\ 
P${^1}\;\;$ & $\nu{^2}\;\;$ & A${^3}\;\;$ & & P$\;\;$ & $\nu \;\;$ & A$\;\;$ & & P$\;\;$ & $\nu \;\;$ & A$\;\;$ & & $\nu \;\;$ & A$\;\;$ & & P$\;\;\;$ & $\nu \;\;\;\;$ & A$\;\;$ & &  $\nu \;\;$ & A$\;\;$ & & P$\;\;$ & $\nu \;\;$ & A$\;\;$  \\
\hline 	         		         			  		                                       	   
      &       &      & &        &         &      & &        &         &      &&          &      & &       &          &      & &         &      & &       &      &      \\ 
\multicolumn{25}{l}{\it Signals Near 650~s} \\
      &       &      & &        &         &      & &        &         &      &&          &      & &        &         &      & &         &      & &       &      &      \\ 
      &       &      & &        &         &      & &        &         &      &&  1535.7  & 7.6  & & 651.01 & 1536.07 & 3.04 & &         &      & &       &      &       \\
      &       &      & & 677.0  & 1477    & 10.0 & &        &         &      &&  1540.8  & 5.6  & & 648.48 & 1542.06 & 5.29 & &         &      & &       &      &       \\
648.1 & 1543  & 14.5 & & 678.0  & 1475    & 9.7  & & 657.0  & 1522    & 12.2 &&  1541.5  & 5.3  & & 648.24 & 1542.64 & 7.71 & &         &      & & 648.6 & 1542 & 17.5 \\
639.6 & 1563  & 3.3  & & 648.5  & 1542    & 6.0  & & 661.4  & 1512    & 7.2  &&  1543.6  & 3.1  & & 647.37 & 1544.70 & 5.50 & & 1544    & 15.6 & & 654.0 & 1529 & 5.0  \\
      &       &      & & 653.6  & 1530    & 4.8  & & 658.3  & 1519    & 5.6  &&  1545.6  & 5.2  & & 646.45 & 1546.90 & 7.28 & & 1548    & 3.6  & &       &      &      \\
      &       &      & &        &         &      & &        &         &      &&  1549.4  & 16.4 & & 645.40 & 1549.42 & 7.84 & &         &      & &       &      &      \\
      &       &      & &        &         &      & &        &         &      &&  1550.7  & 2.9  & & 644.63 & 1551.28 & 2.90 & &         &      & &       &      &      \\
\multicolumn{25}{l}{\it Signals Near 370~s} \\
      &       &      & &        &         &      & &        &         &      &&          &      & &        &         &      & &         &      & &       &      &      \\ 
      &       &      & & 396.7  & 2521    & 8.7  & &        &         &      &&  2644.8  & 4.2  & & 378.10 & 2644.83 & 4.61 & &         &      & &       &      &       \\
377.1 & 2652  & 10.9 & & 387.7  & 2579    & 6.1  & &        &         &      &&  2646.1  & 4.2  & & 377.94 & 2645.90 & 5.39 & &         &      & & 377.4 & 2650 & 10.5  \\
363.1 & 2754  & 3.4  & & 389.1  & 2570    & 5.1  & & 383.6  & 2607    & 9.3  &&  2647.0  & 7.5  & & 377.79 & 2646.96 & 4.35 & & 2654    & 11.6 & & 370.6 & 2698 & 3.9   \\
378.1 & 2645  & 3.1  & & 378.9  & 2639    & 4.5  & & 382.8  & 2612    & 8.4  &&  2661.9  & 4.0  & & 375.67 & 2661.91 & 4.71 & &         &      & & 381.6 & 2621 & 3.2   \\
      &       &      & & 395.3  & 2530    & 4.8  & &        &         &      &&  2662.8  & 7.1  & & 375.52 & 2662.94 & 5.51 & &         &      & &       &      &       \\
      &       &      & &        &         &      & &        &         &      &&  2664.0  & 5.9  & & 375.38 & 2663.98 & 5.51 & &         &      & &       &      &       \\
\multicolumn{25}{l}{\it Signals Near 230~s} \\
      &       &      & &        &         &      & &        &         &      &&          &      & &        &         &      & &         &      & &      &       &       \\
      &       &      & &        &         &      & &        &         &      &&  4237.3  & 6.3  & & 235.99 & 4237.49 & 6.03 & & 4237    & 4.5  & & 236.2& 4233 & 7.8   \\
      &       &      & &        &         &      & &        &         &      &&          &      & & 236.03 & 4236.70 & 1.91 & &         &      & &      &       &      \\
      &       &      & &        &         &      & &        &         &      &&          &      & &        &         &      & &         &      & &      &       &      \\
\multicolumn{25}{l}{\it Low-Frequency Signals} \\
      &       &      & &        &         &      & &        &         &      &&          &      & &        &         &      & &         &      & &      &       &     \\ 
      &       &      & & 2753 & 363     & 9.9  & &        &         &      &&          &      & & 7214.9 & 138.6   & 6.22 & & 127     & 5.0  & & 3823 & 262 & 16.1  \\ 
      &       &      & & 1065 & 939     & 8.2  & & 1178.8 & 848     & 6.5  &&          &      & & 4844.2 & 206.4   & 5.75 & & 206     & 5.2  & & 6413 & 156 & 13.3  \\ 
1105.9 & 904  & 5.8  & & 952  & 1051    & 6.4  & & 1487.4 & 672     & 5.8  &&          &      & & 2844.1 & 351.6   & 5.05 & & 352     & 4.5  & & 2331 & 429 & 7.8   \\ 
      &       &      & & 875  & 1143    & 5.1  & & 793.6  & 1260    & 5.6  &&          &      & & 4118.6 & 242.8   & 4.71 & & 243     & 5.1  & &  967 & 1035& 6.9  \\ 
      &       &      & & 7057 & 142     & 5.1  & & 1351.9 & 7397    & 5.3  &&  292.4   & 3.0  & & 3419.5 & 292.4   & 2.87 & &         &      & &      &       &     \\ 
      &       &      & & 4018 & 2489    & 4.9  & &        &         &      &&  294.2   & 4.0  & & 3399.4 & 294.2   & 2.61 & & 295     & 4.6  & &      &       &     \\ 
\multicolumn{25}{l}{\it Other Signals} \\
      &       &      & &        &         &      & &        &         &      &&          &      & & 288.35  & 3468.05 & 3.04& & 3468    & 3.1  & &      &       &     \\ 
      &       &      & &        &         &      & &        &         &      &&          &      & &        &         &      & &         &      & &      &       &     \\ 
\hline 
\hline
\multicolumn{25}{l}{\footnotesize $^{1}$Period in seconds. $^{2}$Frequency in $\mu$Hz. $^{3}$Amplitude in millimagnitudes.} 
\end{tabular}
\end{sidewaystable*}

\addtocounter{table}{1}

\subsection{The multiplets near 370~s and 650~s}

\begin{figure}
\hspace{2.5mm}\psfig{file=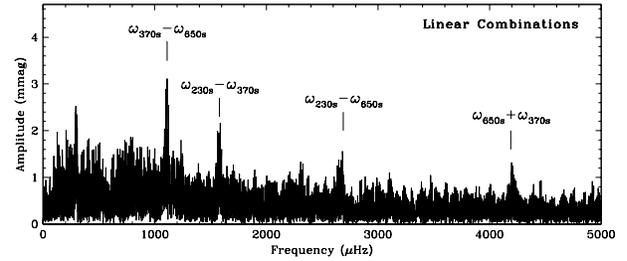,width=8cm}
\caption{Linear combination modes in the residuals of the May\&June98 light-curves after prewhitening with all the components of the 650-s, 370-s and 230-s modes.}\label{fig:lincombs}
\end{figure}

The other principal regions of power in the pulsation spectrum of the May98 dataset -- the structures near 650~s and 370~s -- are a great deal more complex than \two. Fig.~\ref{fig:yukky} shows the 370- and 650-s regions of the May98 and May\&June98 amplitude spectra in detail.

\sthr appears to consist of two triplets, each with components spaced \til1~$\mu$Hz apart (Table 2). One triplet, centred at 2645.90~$\mu$Hz or 377.94~s (indicated by the open arrows in Fig.~\ref{fig:yukky}, which show the three highest-amplitude aliases of the complete triplet), has three closely-spaced components well resolved in both the May98 and May\&June98 FTs. No feature of the spectral window corresponds to the frequency spacing of the triplet components, so we do not believe that they are an observational artifact. \thr's second structure appears to be an unresolved triplet centred at 2662.94~$\mu$Hz or 375.52~s (indicated by the solid arrows in Fig.~\ref{fig:yukky}). Prewhitening this structure produces 3 signals also \til1~$\mu$Hz apart. We fitted these three signals simultaneously to the May98 dataset, the May\&June98 dataset, and two additional datasets containing only the first or the second half of each night. In all four cases, this feature was consistent with a triplet with components \til1~$\mu$Hz apart.

The presence of both a resolved triplet and an unresolved triplet whose components have the same spacing is not inconsistent: if a light-curve is full of substantial gaps (due to daylight or bad weather, etc), then the resolvability of closely-spaced signals whose beat period is approximately equal to the total run length (as is the case here) depends on how the data and the gaps are distributed across their beat cycle. If the data cover regions where the signals are out of phase with each other, the FT is far more likely to resolve the individual signals than if the signals are in phase. We have modelled pairs of triplets made up of closely-spaced sinusoids sampled in the same way as our data, and we are able to reproduce the appearance of \sthr very well.

\six, on the other hand, is a partially-resolved tangle of at least 7 closely- but unevenly-spaced signals (Table 2). It is difficult to know whether these signals represent individual pulsation modes active within the white dwarf, or whether we are seeing the effects of non-stationary behaviour in the pulsation spectrum (i.e., frequency, amplitude or phase modulation). The O-C curves in Fig.~\ref{fig:ominusc} (middle and bottom panels) show noisy modulations, which support our findings that \sthr and \ssix appear to be made up of several different signals.


\subsection{Linear combinations}

\begin{table}
\begin{center}
\caption{Linear combinations.}
\begin{tabular}{cccc} \hline\hline
Combination & Calculated & $\; \;$ Observed $\; \;$ & $\Delta \omega$  \\
            & ($\mu$Hz)  & ($\mu$Hz)  & ($\mu$Hz) \\
\hline
$a_2 + b_3$ & 4196.38 & 4196.47       & 0.09 \\
$a_3 + b_2$ & 4190.60 & 4191.06       & 0.46 \\
$a_4 + b_3$ & 4189.60 & 4189.67 & 0.07 \\
$a_1 + b_4$ & 4209.83 & 4210.64       & 0.81 \\
$a_2 + b_4$ & 4212.36 & 4212.77 & 0.41 \\
$a_1 + b_3$ & 4193.86 & 4193.60       & 0.26 \\
$a_4 + b_1$ & 4187.48 & 4185.46 & 2.02 \\
$a_2 + b_1$ & 4214.26 & 4216.86 & 2.60 \\
$a_3 + b_4$ & 4207.63 & 4208.83 & 1.20 \\
            &         &               &  \\
$b_3 - a_1$ & 1100.06 & 1100.02 & 0.04 \\
$b_3 - a_4$ & 1104.32 & 1104.26 & 0.06 \\
$b_1 - a_1$ & 1097.94 & 1098.35 & 0.41 \\
$b_2 - a_1$ & 1099.01 & 1098.35 & 0.66 \\
$b_1 - a_2$ & 1095.41 & 1095.00       & 0.41 \\
$b_2 - a_3$ & 1101.21 & 1101.74 & 0.53 \\
$b_4 - a_4$ & 1120.29 & 1120.77       & 0.48 \\
$b_4 - a_2$ & 1113.51 & 1113.64       & 0.13 \\
$b_4 - a_3$ & 1118.24 & 1117.62       & 0.62 \\
$b_3 - a_2$ & 1097.54 & 1098.75       & 1.21 \\
            &         &               &      \\
$c - a_3$ & 2692.80 & 2692.54   & 0.26 \\
$c - a_1$ & 2690.60 & 2691.03   & 0.43 \\
$c - a_7$ & 2686.37 & 2686.11         & 0.26 \\
$c - a_4$ & 2694.85 & 2693.24   & 1.61 \\
$c - a_2$ & 2688.07 & 2688.21   & 0.14 \\
          &         &                 &  \\
$c - b_4$ & 1574.56 & 1575.56   & 1.00 \\
$c - b_4$ & 1574.56 & 1574.48         & 0.08 \\
$c - b_2$ & 1591.59 & 1591.49         & 0.10 \\
$c - b_4$ & 1574.56 & 1573.46         & 1.10 \\
$c - b_3$ & 1590.53 & 1589.97   & 0.56 \\
$c - b_1$ & 1592.66 & 1592.26         & 0.40 \\
 \hline\hline
\end{tabular}
$\;$ \\
$a_i =$ the 6 highest-amplitude components at 650~s \\   
$b_i =$ the 6 highest-amplitude components at 370~s \\
$c \; =$ the dominant component of the 230s doublet \\
$\Delta \omega$ = observed -- calculated
\end{center}
\end{table}

Linear combination modes are frequently observed in the pulsation spectra of DAVs. They arise as a result of non-linear mixing of the real eigenmodes: the thickness of the convection zone is modulated by the pulsations and the signal is not only diminished and delayed, but also distorted. This latter distortion gives rise to the power seen at linear combinations (Brickhill 1992).

We have detected four clusters of signals at frequencies corresponding to linear combinations in the May98 and May\&June98 amplitude spectra. In Fig.~\ref{fig:lincombs} we plot the FT of the residuals of the May\&June98 dataset after prewhitening with \six, \sthr  and \two. The amplitudes of these linear combination modes are very low, only a factor of two above the noise level. If they are present in the 1997 and 2001 amplitude spectra, which have greater noise levels, they would therefore not be detectable.

Three of the four linear combination regions lie very close to the dominant signals: the combination (${\omega}_{_{230}}-{\omega}_{_{370}}$) is only \til 40 $\mu$Hz from ${\omega}_{_{650}}$, (${\omega}_{_{230}}-{\omega}_{_{650}}$) is \til 30 $\mu$Hz from ${\omega}_{_{370}}$, and (${\omega}_{_{650}}+{\omega}_{_{370}}$) is \til 40 $\mu$Hz from ${\omega}_{_{250}}$. The proximity of the linear combinations to the dominant signals is unfortunate; at these small separations the window functions partially overlap, which almost certainly contributes to the confusion in ${\omega}_{_{650}}$ and ${\omega}_{_{370}}$.

Our choice of whether a signal is due to a real or combination mode is dictated by their relative amplitudes: the combination mode is defined to have a smaller amplitude than the  constituent real modes. Despite their low signal-to-noise, we discovered 24 signals in the
linear combination regions that corresponded to better than 1 $\mu$Hz
(some to better than 0.1 $\mu$Hz) to the set of 43 combinations
calculated from the highest-amplitude components of \six, \sthr and
\stwo (Table 3). By calculating the fraction of the frequency
axis that is filled by possible linear combinations, we estimate
that only 6 coincidences would be expected by chance alignments.
Thus, we have confidence that the fine structure in \six\
and \thr\ is real: a signal is more likely to be real if it appears in
linear combinations as well (Kleinman et al. 1998).

\section{Discussion}

GW Lib appears similar to the cool DAVs (its pulsation periods are consistent with those expected for non-radial g-modes in single DAVs), which have highly unstable pulsation spectra, with modes disappearing and reappearing on monthly or yearly time-scales, and which therefore require several years -- sometimes decades -- of multisite observing campaigns in order that enough modes are detected for mode identification (and therefore asteroseismology) to be possible. Many more modes will need to be identified before we can begin to unravel GW Lib's pulsation spectrum. But is there anything we can say about its pulsation modes at this early stage?

Already, there are hints that GW Lib's pulsation spectrum is different from those of single DAVs. The frequency splitting is much finer than the multiplet splitting observed in single DAVs, and if caused by rotational splitting, would imply that GW Lib's primary is rotating an order of magnitude more slowly than single DAVs, which for a CV primary is improbable. In addition, if we consider \thr's two triplets to be modes of same $l$ (because the spacing of their components is the same) then the separation between modes is too small. From Bradley's (1996) models, the minimum average separation for DAV modes of $l=1$ is \til34~s, and \til20~s for $l=2$ modes, while the two triplets within \sthr are separated by only \til2~s. Two triplets of {\em different} $l$ could have a small separation, but then their component spacings would be different.

The April 1997 amplitude spectrum (Fig.~\ref{fig:details}) shows (at least) three `modes' near 370~s (each mode appears to consist of more than one structure, so may therefore be composed of two or more multiplets, like \thr). The right-hand April97 mode (at 378.9) differs by only \til6~$\mu$Hz from the left-hand triplet in \thr, so -- given the poorer resolution of the April97 dataset -- these may be the same mode. This still gives us four, perhaps more, modes lying within a 20-s interval. Even if we imagine them to be modes of alternating $l=1$ and $l=2$, they are still too close together.

These awkward period spacings, both the fine splitting within apparent modes and the mode-to-mode spacing, show that it is going to be difficult to understand GW Lib's pulsation spectrum using the framework appropriate to single, non-accreting DAVs. It may be that the assumptions implicit in the above paragraphs (i.e., that the modes are all of high radial order, and that only modes of $l=1$ and $l=2$ are present) don't hold for accreting DAVs. It may be that in GW Lib we are observing, for the first time in a DAV, modes of $l>2$. In \S1.2.2 we point out that geometric cancellation may work differently for stars with hot equatorial bands, allowing modes of higher $l$ to be visible. A possible way to identify whether GW Lib's modes are of $l=3$ or higher would be to obtain simultaneous UV and optical light-curves; higher-order modes have higher UV amplitudes because limb-darkening helps to negate the effects of geometric cancellation at these wavelengths (Robinson et al. 1995; Kepler et al. 2000). 
In the case of GW Lib an appropriate model for the fractional 
pulsation amplitudes would have to be developed, taking into account 
possible accretion-related differences in surface temperature 
distribution and their effect on limb-darkening and geometric 
cancellation, before this mode-identification technique could be 
applied.

If we are to understand GW Lib's pulsation spectrum, many more pulsation modes will need to be identified, which will require a substantial observational effort. In addition, the behaviour of $g$-modes in accreting, rapidly rotating white dwarfs with latitude-dependent temperature, angular momentum and chemical composition structures in their outer layers will need to be modelled, as will the effects of DN outbursts and superoutbursts on the stars' eigenfrequencies.

\section*{Acknowledgments}

Based on observations obtained at the Southern African Astronomical Observatory (South Africa), Mt John Observatory (New Zealand), Mt Stromlo \& Siding Spring Observatory (Australia), McDonald Observatory (USA), Kitt Peak Observatory (USA) and CTIO (Chile).

LvZ would like to thank Scott Kleinman, Jurek Krzesinski and Don Kurtz for helpful comments and discussions.

\label{lastpage}


\begin{thebibliography}{}

\bibitem{}
Augusteijn T., van der Hooft F., de Jong J.A., van Paradijs J. 1996, A\&A, 311, 889
\bibitem{}
Bailey J. 1979, MNRAS, 189, 41
\bibitem{}
Bradley P.A., 2001, ApJ, 552, 326
\bibitem{}
Bradley P.A., 1996, ApJ, 468, 350
\bibitem{}
Brickhill A.J., 1992, MNRAS, 259, 519
\bibitem{}
Chabrier G., Brassard P., Fontaine G., Saumon D. 2000, ApJ, 543, 216
\bibitem{}
Deeming T.J., 1975, Astr.\&Sp.Sci, 36, 
\bibitem{}
Duerbeck H.W., Seitter, W.C., 1987, ApSS, 131, 467
\bibitem{}
Gonz\'{a}lez L.E., 1983, IAU Circ 3854
\bibitem{}
Howell S.B., Szkody P., Cannizzo J.K. 1995, ApJ, 439, 337
\bibitem{}
Kepler S.O., et al. 2000, Baltic Astr., 9, 59
\bibitem{}
Kleinman S.J., 1998, private communication
\bibitem{}
Kleinman S.J., et al., 1998, ApJ, 495, 424
\bibitem{}
Koester D., Holberg J. B. 2001, 12th European Workshop on White Dwarfs, ASP Conference Proceedings, 226, eds. J. L. Provencal, H. L. Shipman, J. MacDonald, and S. Goodchild, San Francisco, p.299
\bibitem{}
Koester D., Dreizler S., Weidemann V., Allard N.F. 1998, A\&A, 338, 612
\bibitem{}
Kurtz D.W., 1985, MNRAS 213, 773
\bibitem{}
Long K.S., et al., 1994, ApJ, 424, L49
\bibitem{}
O'Donoghue D., Chen A., Marang F., Mittaz J. P. D., Winkler H., Warner B. 1991, MNRAS, 250, 363
\bibitem{}
O'Donoghue D. 1995, Baltic Astr., 4, 517
\bibitem{}
Ringwald, F.~A., Naylor, T. and Mukai, K. 1996, MNRAS, 281, 192
\bibitem{}
Robinson E.L., et al. 1995, ApJ, 438, 908
\bibitem{}
Schou J., et al. 1998, ApJ, 505, 390
\bibitem{}
Sion E.M., Szkody P., 1990, IAU Colloq. No. 122, p. 59
\bibitem{}
Sion E.M., Urban J., Lyons K. 2001, 12th European Workshop on White Dwarfs, ASP Conference Proceedings, 226, eds. J. L. Provencal, H. L. Shipman, J. MacDonald, and S. Goodchild, San Francisco, p.205
\bibitem{}
Solheim J.-E., et al. 1998, A\&A, 332, 939
\bibitem{}
Sparks W.M., Sion E.M., Starrfield S.G., Austin S., 1992, ASP Conf. Ser. 29, 167
\bibitem{}
Sparks W.M., et al., 1993, in Cataclysmic Variables and Related Physics, eds. O. Regev \& G. Shaviv, Inst. Phys. Publ., Bristol, P. 96
\bibitem{}
Szkody P., Desai V., Hoard D.~W. 2000, AJ, 119, 365
\bibitem{}
Szkody P., Gansicke B., Sion E.M., Howell S. 2002, ApJ, 575, L79 
\bibitem{}
Thorstensen J.R., Patterson J.O., Kemp J. and Vennes S. 2002, PASP, in press, astro-ph/0206426
\bibitem{}
Townsley D.M., Bildsten L. 2002, ASP Conf. Proc., 261, eds. B.T. G\"{a}nsicke, K. Beuermann, K. Reinsch, p. 31
\bibitem{}
van Zyl et al, 2000, Baltic Astr., 9, 231 (Proceedings of the Fifth Whole Earth Telescope Workshop)
\bibitem{}
Voges W. et al. 2000, IAU Circ., 7432
\bibitem{}
Vorontsov S.V., 1993, ASP Conf. Ser 40, eds W.W. Weiss \& A. Baglin, p 497
\bibitem{}
Warner B. 1995, Cataclysmic Variable Stars, Cambridge University Press
\bibitem{}
Winget D.E., et al., 1991, ApJ, 378, 326
\bibitem{}
Winget D.E., et al., 1994, ApJ, 430, 839
\bibitem{}
Wood M.A. 1992, ApJ, 386, 539

\end{thebibliography}
\end{document}